\documentstyle[aps,preprint,tighten,floats,psfig,epsf,rotate]{revtex}

\newcommand{\be}{\begin{equation}}
\newcommand{\ee}{\end{equation}}

\draft
\preprint{\vbox{
 \hfill \rm \null \hfill
 \hbox{ADP-01-39/T471} \\
 \hfill \rm \null \hfill
 \hbox{JLAB-THY-01-32}
}}

\begin{document}

\title{\bf Hadron Masses From Novel Fat-Link Fermion Actions }

\author{
J.~M.~Zanotti,
S.~Bilson-Thompson,
F.~D.~R.~Bonnet,
P.~D.~Coddington,
D.~B.~Leinweber,
A.~G.~Williams and
J.~B.~Zhang}

\address{ \vspace*{3mm}
        {CSSM Lattice Collaboration, } \\
	Special Research Center for the
	Subatomic Structure of Matter, and		\\
	Department of Physics and Mathematical Physics,
	University of Adelaide, 5005, Australia}
\author{W.~Melnitchouk}
\address{
	Special Research Center for the
	Subatomic Structure of Matter, and		\\
	Department of Physics and Mathematical Physics,
	University of Adelaide, 5005, Australia,	\\
	and Jefferson Lab, 12000 Jefferson Avenue,
	Newport News, VA 23606}
\author{F.~X.~Lee}
\address{
	Center for Nuclear Studies, Department of Physics,	\\
	The George Washington University, Washington, D.C. 20052,\\
	and Jefferson Lab, 12000 Jefferson Avenue,
	Newport News, VA 23606}
 
\maketitle

\begin{abstract}
The hadron mass spectrum is calculated in lattice QCD using a novel 
fat-link clover fermion action in which only the irrelevant operators
in the fermion action are constructed using smeared links.
The simulations are performed on a $16^3 \times 32$ lattice with a
lattice spacing of $a=0.125$ fm.
We compare actions with $n=4$ and 12 smearing sweeps with a smearing
fraction of 0.7. The $n=4$ Fat-Link Irrelevant Clover (FLIC)
action provides scaling which is superior to mean-field improvement,
and offers advantages over nonperturbative ${\cal O}(a)$ improvement, 
including a reduced exceptional configuration problem.
\end{abstract}

\vspace{3mm}
PACS number(s): 11.15.Ha, 12.38.Gc, 12.38.Aw

\newpage

\section{Introduction}

The origin of the masses of light hadrons represents one of the most
fundamental challenges to the theory of strong interactions, Quantum
Chromodynamics (QCD).
Despite the universal acceptance of QCD as the basis from which to derive
hadronic properties, there has been slow progress in understanding the
generation of hadron mass from first principles.
Solving the problem of the hadronic mass spectrum would allow considerable
improvement in our understanding of the nonperturbative nature of QCD.
The only available method at present to derive hadron masses directly
from QCD is a numerical calculation on the lattice.
In the last few years impressive progress has been made both in computer
hardware and in developing more efficient algorithms, bringing realistic
simulations of hadronic observables with sufficiently large volumes, small 
quark masses and fine enough lattices within reach.

The high computational cost required to perform accurate lattice
calculations at small lattice spacings, however, has led to
increased interest in quark action improvement.
In order to avoid the famous doubling problem, Wilson \cite{Wilson}
originally introduced an irrelevant (energy) dimension-five operator
(the ``Wilson term'') to the standard ``naive'' lattice fermion action,
which explicitly breaks chiral symmetry at ${\cal O}(a)$.
To extrapolate reliably to the continuum, simulations must be performed
on fine lattices, which are therefore very computationally expensive.
The scaling properties of the Wilson action at finite $a$ can be
improved by introducing any number of irrelevant operators of increasing
dimension which vanish in the continuum limit.

The Sheikholeslami-Wohlert (clover) action \cite{Sheikholeslami:1985ij}
introduces an additional irrelevant dimension-five operator to the
standard Wilson \cite{Wilson} quark action, 
\be
S_{\rm SW} = S_{\rm W} - \frac{i a C_{\rm SW} r}{4}\
	 \bar{\psi}(x)\sigma_{\mu\nu}F_{\mu\nu}\psi(x)\ ,
\label{clover}
\ee
where $S_{\rm W}$ is the standard Wilson action,
\be
S_{\rm W} = \bar{\psi}(x)\left[ \sum_{\mu} \left( \gamma_{\mu}\, \nabla_{\mu}
    - \frac{1}{2} r a \Delta_{\mu} \right) +m\right] \psi(x)\ ,
\ee
$\nabla_{\mu}$ and $\Delta_{\mu}$ are the standard covariant first
and second order lattice derivatives, 
\begin{eqnarray*}
\nabla_\mu \psi(x) &=& {1\over2a}  \left[ U_\mu(x)\psi(x+\mu)-U_\mu^\dagger(x-\mu)\psi(x-\mu)\right], \\
\Delta_\mu \psi(x) &=& {1\over a^2}\left[ U_\mu(x)\psi(x+\mu)+U_\mu^\dagger(x-\mu)\psi(x-\mu)-2\psi(x)\right],
\end{eqnarray*}
and $C_{\scriptstyle {\rm SW}}$ is the clover coefficient which can be tuned to
remove ${\cal O}(a)$ artifacts,
\begin{eqnarray}
C_{\rm SW} &=&
\left\{
\begin{tabular}{l}
	1  \, \mbox{\rm{\ \ \ at tree-level\, ,}}	\\
	$1/u_0^3$\, \mbox{\rm{mean-field improved\, ,}}
\end{tabular}
\right.
\end{eqnarray}
with $u_0$ the tadpole improvement factor which corrects for the
quantum renormalization of the operators.
Nonperturbative (NP) ${\cal O}(a)$ improvement \cite{Luscher:1996sc}
tunes $C_{\scriptstyle {\rm SW}}$ to all powers in $g^2$ and displays excellent
scaling, as shown by Edwards {\it et al}. \cite{Edwards:1998nh}, who studied
%
%
the scaling properties of the nucleon and vector meson masses for various
lattice spacings (see also Section~\ref{discussion} below).
In particular, the linear behavior of the NP-improved clover actions,
when plotted against $a^2$, demonstrates that ${\cal O}(a)$
errors are removed.
It was also found in Ref.~\cite{Edwards:1998nh} that a linear
extrapolation of the mean-field improved data fails,
indicating that ${\cal O}(a)$ errors are still present.

A drawback to the clover action, however, is the associated problem of
exceptional configurations, where the quark propagator encounters
singular behavior as the quark mass becomes small.
In practice, this prevents the use of coarse lattices
($\beta \agt 5.7 \sim a \alt 0.18$~fm)
\cite{Bardeen:1998gv,DeGrand:1998jq}.
Furthermore, 
%
the plaquette version of $F_{\mu\nu}$, which is commonly used in
Eq.~(\ref{clover}), has large ${\cal O}(a^2)$ errors, which can lead to
errors of the order of $10 \%$ in the topological charge even on very
smooth configurations \cite{Bonnet:2000dc}.

The idea of using fat links in fermion actions
was first explored by the MIT group \cite{Chu:1994vi}
and more recently has been studied by DeGrand {\it et al}.
\cite{DeGrand:1998jq,DeGrand:1999gp,SDDH}, who showed that the exceptional
configuration problem can be overcome by using a fat-link (FL) clover
action.
Moreover, the renormalization of the coefficients of action improvement
terms is small.
In principle it is acceptable to smear the links of the relevant
operators. The symmetry of the APE smearing process ensures that effects are
${\cal O}(a^2)$. The factors multiplying the link and staple ensure the leading order
term is $e^{iagA_\mu}$, an element of SU(3). Issues of projecting the
smeared links to SU(3) are ${\cal O}(a^2)$ effects and therefore correspond
to irrelevant operators \cite{BD}.
However, the net effect of APE smearing the links of
the relevant operators is to remove
%
%
gluon interactions at the scale of the cutoff.
While this has some tremendous benefits, the short-distance quark
interactions are lost. As a result decay constants, which are sensitive
to the wave function at the origin, are suppressed.

A possible solution to this is to work with two sets of links in the fermion
action. 
In the relevant dimension-four operators, one works with the untouched
links generated via Monte Carlo methods, while the smeared fat links
are introduced only in the higher dimension irrelevant operators. 
The effect this has on decay constants is under investigation and
will be reported elsewhere.

In this paper we present the first results of simulations of the
spectrum of light mesons and baryons using this variation on the clover
action.
%
%
In particular, we will start with the standard clover action and replace
the links in the irrelevant operators with APE smeared \cite{ape}, or fat
links.
We shall refer to this action as the Fat-Link Irrelevant Clover (FLIC)
action.
Although the idea of using fat links only in the irrelevant operators of
the fermion action was developed here independently, suggestions have
appeared previously \cite{neub}.
To the best of our knowledge, this is the first report of lattice QCD
calculations using this novel fermion action.

In Section~\ref{simulations} we describe the gauge action used in our 
lattice simulations, while Section~\ref{FLinks} contains the procedure
for creating the FLIC fermion action.
Our results are presented in Section \ref{discussion}, and finally in
Section \ref{conclusion} we draw some conclusions and discuss future work.

\section{The Gauge Action}
\label{simulations}

The simulations are performed using a mean-field improved, plaquette
plus rectangle, gauge action on a $16^3 \times 32$ lattice at
$\beta = 6/g^2 = 4.60$, providing a lattice spacing $a=0.125(2)$~fm
determined from the string tension with $\sqrt\sigma=440$~MeV.
The tree-level ${\cal O}(a^2)$--Symanzik-improved gauge action
\cite{Symanzik:1983dc} is defined as
\be
S_{\rm G} = \frac{5\beta}{3}
      \sum_{\rm{sq}}{\cal R}e\ {\rm{tr}}(1-U_{\rm{sq}}(x))
    - \frac{\beta}{12u_{0}^2}
      \sum_{\rm{rect}}{\cal R}e\ {\rm{tr}}(1-U_{\rm{rect}}(x))\ ,
\label{gaugeaction}
\ee
where the operators $U_{\rm{sq}}(x)$ and $U_{\rm{rect}}(x)$ are
defined as
\begin{mathletters}
\begin{eqnarray}
U_{\rm{sq}}(x)
&=& U_{\mu}(x) U_{\nu}(x+\hat{\mu})
    U^{\dagger}_{\mu}(x+\hat{\nu}) U^\dagger_{\nu}(x)\ ,	\\
U_{\rm{rect}}(x)
&=& U_{\mu}(x) U_{\nu}(x+\hat{\mu})
    U_{\nu}(x+\hat{\nu}+\hat{\mu})			\nonumber\\
& &\times U^{\dagger}_{\mu}(x+2\hat{\nu})
	 U^{\dagger}_{\nu}(x+\hat{\nu})
	 U^\dagger_{\nu}(x) 				\nonumber\\
&+& U_{\mu}(x) U_{\mu}(x+\hat{\mu})
    U_{\nu}(x+2\hat{\mu})				\nonumber\\
& &\times U^{\dagger}_{\mu}(x+\hat{\mu}+\hat{\nu})
	 U^{\dagger}_{\mu}(x+\hat{\nu})U^\dagger_{\nu}(x)\ .
\label{actioneqn}
\end{eqnarray}
\end{mathletters}%
The link product $U_{\rm{rect}}(x)$ denotes the rectangular $1\times2$
and $2\times1$ plaquettes, and for the tadpole improvement factor we
employ the plaquette measure
\be
u_0 = \left( \frac{1}{3}{\cal R}e \, {\rm{tr}}\langle U_{\rm{sq}}\rangle
      \right)^{1/4}\ .
\label{uzero}
\ee
Gauge configurations are generated using the Cabibbo-Marinari
pseudoheat-bath algorithm with three diagonal SU(2) subgroups looped
over twice.
Simulations are performed using a parallel algorithm with appropriate
link partitioning \cite{Bonnet:2001db}.
A total of 50 configurations are used in this analysis, and the error
analysis is performed by a third-order, single-elimination jackknife,
with the $\chi^2$ per degree of freedom ($N_{\rm DF}$) obtained via
covariance matrix fits.

\section{Fat-Link Irrelevant Fermion Action}
\label{FLinks}

Fat links \cite{DeGrand:1998jq,DeGrand:1999gp} are created by averaging
or smearing links on the lattice with their nearest neighbours in a
gauge covariant manner (APE smearing).
The smearing procedure \cite{ape} replaces a link, $U_{\mu}(x)$, with a
sum of the link and $\alpha$ times its staples
\begin{eqnarray}
U_{\mu}(x)\ \rightarrow\ U_{\mu}'(x)\ =\
(1-\alpha) U_{\mu}(x)
+ \frac{\alpha}{6}\sum_{\nu=1 \atop \nu\neq\mu}^{4}
  \Big[	U_{\nu}(x)
	U_{\mu}(x+\nu a)
	U_{\nu}^{\dag}(x+\mu a)				\nonumber \\
\mbox{} 
      + U_{\nu}^{\dag}(x-\nu a)
	U_{\mu}(x-\nu a)
	U_{\nu}(x-\nu a +\mu a)
  \Big] \,,
\end{eqnarray} 
followed by projection back to SU(3). We select the unitary matrix
$U_{\mu}^{\rm FL}$ which maximizes
$$
{\cal R}e \, {\rm{tr}}(U_{\mu}^{\rm FL}\, U_{\mu}'^{\dagger})\, ,
$$
by iterating over the three diagonal SU(2) subgroups of SU(3).
We repeat the combined procedure of smearing and projection
$n$ times.
We create our fat links by setting $\alpha = 0.7$ and comparing $n=4$ and
12 smearing sweeps.
The mean-field improved FLIC action now becomes
\be
S_{\rm SW}^{\rm FL}
= S_{\rm W}^{\rm FL} - \frac{iC_{\rm SW} \kappa r}{2(u_{0}^{\rm FL})^4}\
	     \bar{\psi}(x)\sigma_{\mu\nu}F_{\mu\nu}\psi(x)\ ,
\label{FLIC}
\ee
where $F_{\mu\nu}$ is constructed using fat links, $u_{0}^{\rm FL}$ is
calculated in an analogous way to Eq.~(\ref{uzero}), and where the
mean-field improved Fat-Link Irrelevant Wilson action is
\begin{eqnarray}
S_{\rm W}^{\rm FL}
 =  \sum_x \bar{\psi}(x)\psi(x) 
&+& \kappa \sum_{x,\mu} \bar{\psi}(x)
    \bigg[ \gamma_{\mu}
      \bigg( \frac{U_{\mu}(x)}{u_0} \psi(x+\hat{\mu})
	- \frac{U^{\dagger}_{\mu}(x-\hat{\mu})}{u_0} \psi(x-\hat{\mu})
      \bigg)						\nonumber\\
&-& r \bigg(
	  \frac{U_{\mu}^{\rm FL}(x)}{u_0^{\rm  FL}} \psi(x+\hat{\mu})
	+ \frac{U^{{\rm FL}\dagger}_{\mu}(x-\hat{\mu})}{u_0^{\rm FL}}
	  \psi(x-\hat{\mu})
      \bigg)
    \bigg]\ ,
\label{MFIW}
\end{eqnarray}
with $\kappa = 1/(2m+8r)$. We take the standard value $r=1$.
Our notation uses the Pauli representation of the Dirac $\gamma$-matrices
defined in Appendix B of Sakurai \cite{Sakurai}. In particular, the 
$\gamma$-matrices are hermitian and
$\sigma_{\mu\nu} = [\gamma_{\mu},\ \gamma_{\nu}]/(2i)$.

As reported in Table~\ref{meanlink}, the mean-field improvement parameter
for the fat links is very close to 1.
Hence, the mean-field improved coefficient for $C_{\rm SW}$ is expected
to be adequate{\footnote{Our experience with topological charge operators
suggests that it is advantageous to include $u_0$ factors, even as they
approach 1.}}.
In addition, actions with many irrelevant operators ({\it e.g.} the
D$_{234}$ action) can now be handled with confidence as tree-level
knowledge of the improvement coefficients should be sufficient.
Another advantage is that one can now use highly improved definitions
of $F_{\mu\nu}$ (involving terms up to $u_0^{12})$, which give impressive
near-integer results for the topological charge \cite{sbilson}.

In particular, we employ an ${\cal O}(a^4)$ improved definition of
$F_{\mu\nu}$ in which the standard clover-sum of four $1 \times 1$ Wilson
loops lying in the $\mu ,\nu$ plane is combined with $2 \times 2$ and $3
\times 3$ Wilson loop clovers.
Bilson-Thompson {\it et al.} \cite{sbilson} find
\be
F_{\mu\nu} = {-i\over{8}} \left[\left( {3\over{2}}W^{1 \times 1}-{3\over{20u_0^4}}W^{2 \times 2}
+{1\over{90u_0^8}}W^{3 \times 3}\right) - {\rm h.c.}\right]_{\rm Traceless}\ 
\ee
where $W^{n \times n}$ is the clover-sum of four $n \times n$ Wilson loops
and $F_{\mu\nu}$ is made traceless by subtracting $1/3$ of the trace from
each diagonal element of the $3 \times 3$ color matrix.
This definition reproduces the continuum limit with ${\cal O}(a^6)$
errors.
On approximately self-dual configurations, this operator produces integer 
topological charge to better than 4 parts in $10^4$.

\begin{table}
\begin{center}
\begin{tabular}{ccc}
$n$ & $u^{\rm FL}_0$ & $(u^{\rm FL}_0)^4$ \\ \hline
    0  &  0.88894473 & 0.62445197 \\
    4  &  0.99658530 & 0.98641100 \\
    12 &  0.99927343 & 0.99709689
\end{tabular}
\vspace*{0.5cm}
\caption{The value of the mean link for different numbers of APE smearing
	sweeps, $n$, at $\alpha = 0.7$. \label{meanlink}}
\end{center}
\end{table}


Work by DeForcrand {\it et al}. \cite{deForcrand:1997sq} suggests that 7
cooling sweeps are required to approach topological charge within 1$\%$ of
integer value.
This is approximately 16 APE smearing sweeps at $\alpha = 0.7$
\cite{Bonnet:2001rc}.
However, achieving integer topological charge is not necessary for the
purposes of studying hadron masses, as has been well established.
To reach integer topological charge, even with improved definitions of
the topological charge operator, requires significant smoothing and
associated loss of short-distance information.
Instead, we regard this as an upper limit on the number of smearing
sweeps.

Using unimproved gauge fields and an unimproved topological charge
operator, Bonnet {\it et al}. \cite{Bonnet:2000dc} found that the
topological charge settles down after about 10 sweeps of APE smearing at
$\alpha=0.7$.
Consequently, we create fat links with APE smearing parameters $n=12$ and
$\alpha=0.7$.
This corresponds to $\sim 2.5$ times the smearing used in
Refs.~\cite{DeGrand:1998jq,DeGrand:1999gp}.
Further investigation reveals that improved gauge fields with a small
lattice spacing ($a=0.125$~fm) are smooth after only 4 sweeps.
Hence, we perform calculations with 4 sweeps of smearing at $\alpha=0.7$
and consider $n=12$ as a second reference. 
Table~\ref{meanlink} lists the values of $u_0^{\rm FL}$ for $n=0$, 4 and
12 smearing sweeps.

We also compare our results with the standard Mean-Field Improved Clover
(MFIC) action. We mean-field improve as defined in Eqs.~\ref{FLIC} and
\ref{MFIW} but with thin links throughout. The standard Wilson-loop
definition of $F_{\mu\nu}$ is used.

A fixed boundary condition is used for the fermions by setting
\be
U_t (\vec{x},nt) = 0 \qquad {\rm and} \qquad U_t^{\rm FL} (\vec{x},nt) = 0 
\qquad \forall\ \vec{x}\ 
\ee
in the hopping terms of the fermion action.
The fermion source is centered at the space-time location
{$(x,y,z,t) = (1,1,1,3)$}, which allows for two steps backward in time
without loss of signal. 
Gauge-invariant gaussian smearing in the spatial dimensions is applied at 
the source to increase the overlap of the interpolating operators with the
ground states.


\begin{figure}[t]
\begin{center}
\epsfysize=10.8truecm
\leavevmode
\rotate[l]{\epsfbox{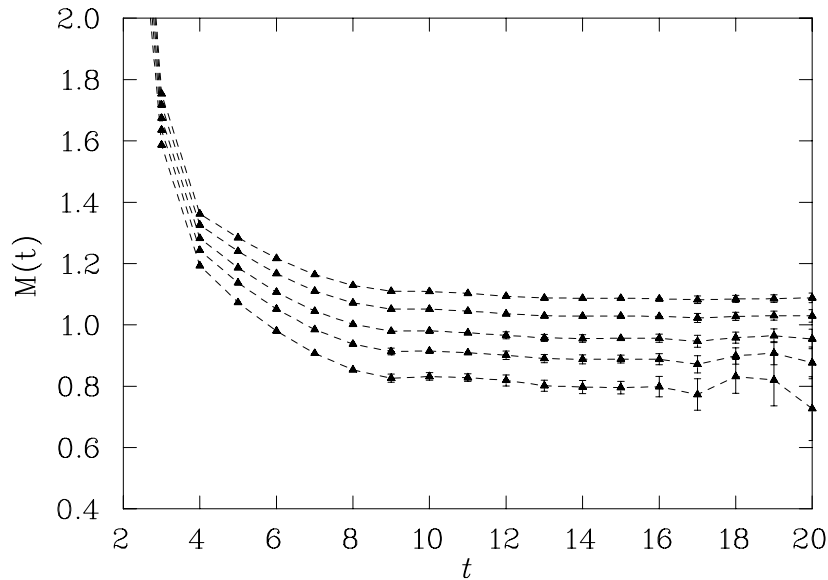} }
\vspace*{0.5cm}
\caption{Effective mass plot for the nucleon for the FLIC action with 4
	sweeps of smearing at $\alpha = 0.7$ from 200 configurations.
	The five sets of points correspond to the $\kappa$ values listed
	in Table~\protect\ref{masses}, with $\kappa$ increasing from top
	down.}
\label{rho}
\end{center}
\end{figure}

\section{Results}
\label{discussion}

Hadron masses are extracted from the Euclidean time dependence of the
calculated two-point correlation functions. For baryons, the correlation
functions are given by
\be
\langle G(t;\vec{p},\Gamma)\rangle
= \sum_x e^{-i\, \vec p\cdot \vec x}\ \Gamma^{\beta\alpha}
\langle\Omega|T[\chi^{\alpha}(x)\bar{\chi}^{\beta}(0)]|\Omega\rangle\ ,
\ee
where $\chi$ are standard baryon interpolating fields, $\Omega$
represents the QCD vacuum, $\Gamma$ is a $4\times4$ matrix in Dirac
space, and $\alpha$, $\beta$ are Dirac indices.
At large Euclidean times one has
\be
\langle G(t;\vec{p},\Gamma)\rangle
\simeq \frac{Z^2}{2 E_p}\ e^{-E_p t}\
{\rm tr}\left[ \Gamma(-i\gamma\cdot p+M)\right]\ ,
\ee
where $Z$ represents the coupling strength of $\chi(0)$ to the baryon,
and $E_p = ({\vec p}^{\, 2} + M^2 )^{1/2}$ is the energy.
Selecting ${\vec p}=0$ and $\Gamma = (1+\gamma_4)/4$, the effective
baryon mass is then given by
\be
M(t+1/2) = \log [G(t)] - \log[G(t+1)]\ .
\ee
Meson masses are determined via analogous standard procedures.
The critical value of $\kappa$, $\kappa_{c}$, is determined by linearly
extrapolating $m_{\pi}^2$ as a function of $m_q$ to zero. 
The quark masses are defined by
$m_q = \left( 1/\kappa - 1/\kappa_{c} \right)/(2a)$,
and the strange quark mass was taken to be the second heaviest quark mass
in each case.

Figure~\ref{rho} shows the nucleon effective mass plot for the FLIC
action when 4 APE smearing sweeps at $\alpha=0.7$ are performed on the
fat links (``FLIC4'').
The effective mass plots for the other hadrons are similar, and all
display acceptable plateau behavior.
Good values of $\chi^2 / N_{\rm DF}$ are obtained for many different
time-fitting intervals as long as one fits after time slice 8.
All fits for this action are therefore performed on time slices 9 through
14.
For the Wilson action and the FLIC action with $n=12$ (``FLIC12'') the
fitting regimes used are 9-13 and 9-14, respectively. 

\begin{table}[t]
\begin{center}
  \begin{tabular}{ccccc}
	$\kappa$ &
	$m_{\pi}\, a$ & $m_{\rho}\, a$ &
	$m_{\rm N}\, a$ & $m_{\Delta}\, a$	\\ \hline
    0.1260 & 0.5797(23) & 0.7278(39) & 1.0995(58)  & 1.1869(104) \\
    0.1266 & 0.5331(24) & 0.6951(45) & 1.0419(64)  & 1.1387(121) \\
    0.1273 & 0.4744(27) & 0.6565(54) & 0.9709(72)  & 1.0816(152) \\
    0.1279 & 0.4185(30) & 0.6229(65) & 0.9055(82)  & 1.0310(194) \\
    0.1286 & 0.3429(37) & 0.5843(97) & 0.8220(102) & 0.9703(286) \\
\end{tabular}
\vspace*{0.5cm}
\caption{Values of $\kappa$ and the corresponding $\pi,\, \rho,\,
	{\rm N}$ and $\Delta$ masses for the FLIC action with 4 sweeps
	of smearing at $\alpha=0.7$. The value for $\kappa_{\rm cr}$ is
        provided in Table~\protect\ref{kappa}. 
	A string tension analysis provides $a=0.125(2)$ fm for
	$\sqrt\sigma=440$~MeV.\label{masses}}
\end{center}
\end{table}

\begin{table}
\begin{center}
  \begin{tabular}{ccccc}
	& Wilson & FLIC12 & FLIC4 & MFIC  	\\ \hline
    $\kappa_1$ & 0.1346 & 0.1286 & 0.1260 & 0.1196 \\
    $\kappa_2$ & 0.1353 & 0.1292 & 0.1266 & 0.1201 \\
    $\kappa_3$ & 0.1360 & 0.1299 & 0.1273 & 0.1206 \\
    $\kappa_4$ & 0.1367 & 0.1305 & 0.1279 & 0.1211 \\
    $\kappa_5$ & 0.1374 & 0.1312 & 0.1286 & 0.1216 \\ \hline
    $\kappa_{\rm cr}$ & 0.1390 & 0.1328 & 0.1300 & 0.1226

\end{tabular}
\vspace*{0.5cm}
\caption{Values of $\kappa$ and $\kappa_{\rm cr}$ for the four different actions.
        \label{kappa}}

\end{center}
\end{table}

\begin{figure}[t]
\begin{center}
\epsfysize=11.3truecm
\leavevmode
\rotate[l]{\epsfbox{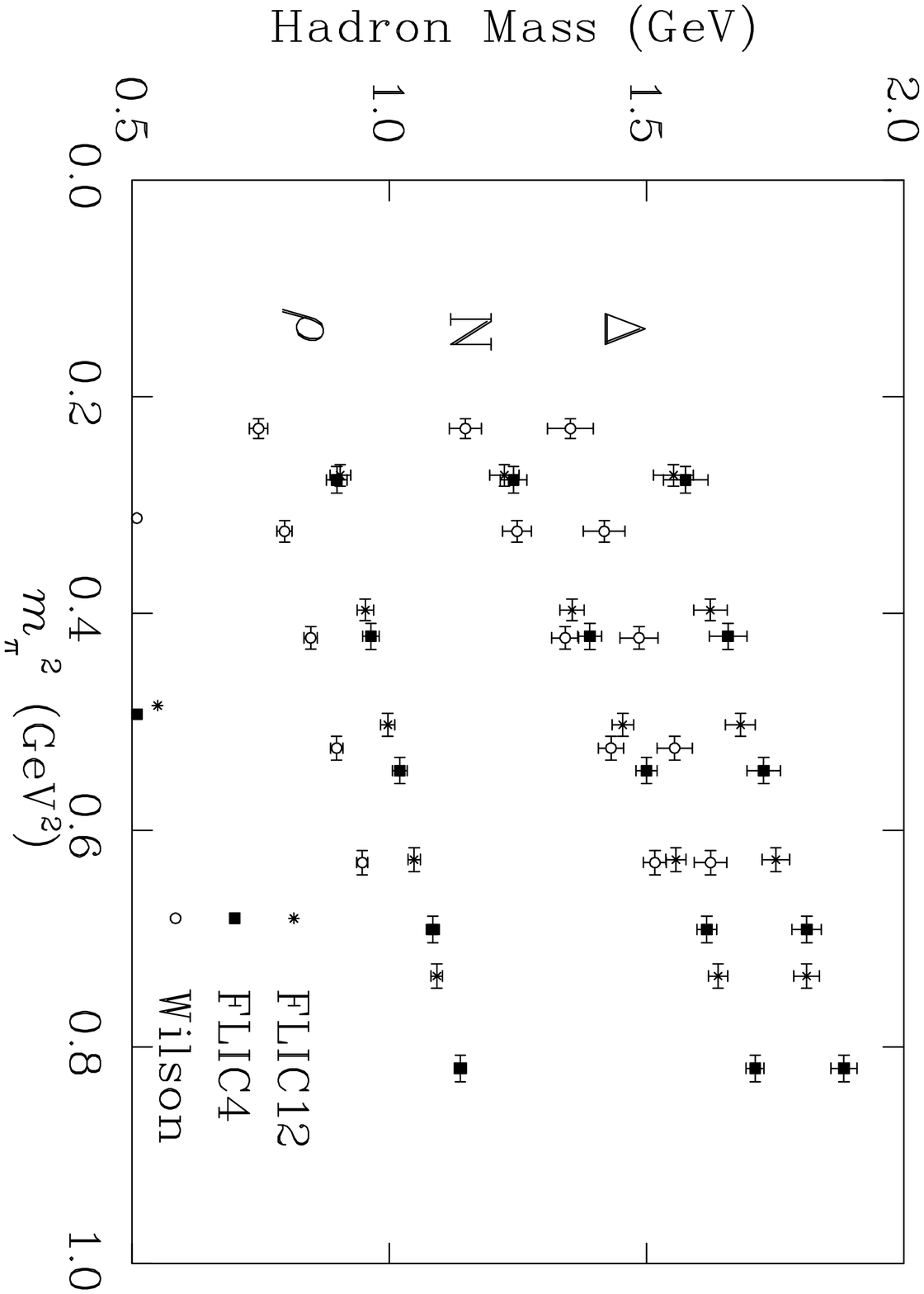} }
\vspace*{0.5cm}
\caption{Masses of the nucleon, $\Delta$ and $\rho$ meson versus
	$m_{\pi}^2$ for the FLIC4, FLIC12 and Wilson actions.}
\label{Mvsmpi2}
\end{center}
\end{figure}

The values of $\kappa$ used in the simulations for all quark actions
are give in Table~\ref{kappa}. We have also provided the values of
$\kappa_{\rm cr}$ for these fermion actions when using our mean-field 
improved, plaquette plus rectangle, gauge action at $\beta = 4.60$. 
We have mean-field improved our fermion actions so we expect
the values for $\kappa_{\rm cr}$ to be close to the tree-level value of 0.125.
Improved chiral properties are seen for the FLIC and MFIC actions, with FLIC4
performing better than FLIC12.

The behavior of the $\rho$, nucleon and $\Delta$ masses as a function of
squared pion mass is shown in Fig.~\ref{Mvsmpi2} for the various actions.
The first feature to note is the excellent agreement between the FLIC4
and FLIC12 actions.
On the other hand, the Wilson action appears to lie somewhat low in
comparison.
It is also reassuring that all actions give the correct mass ordering in
the spectrum.
The value of the squared pion mass at $m_{\pi}/m_{\rho} = 0.7$ is plotted
on the abscissa for the three actions as a reference point.
This point is chosen in order to allow comparison of different results by
interpolating them to a common value of $m_{\pi}/m_{\rho} = 0.7$, rather
than extrapolating them to smaller quark masses, which is subject to
larger systematic and statistical uncertainties.

\begin{figure}[t]
\begin{center}
\epsfysize=11.3truecm
\leavevmode
\rotate[l]{\epsfbox{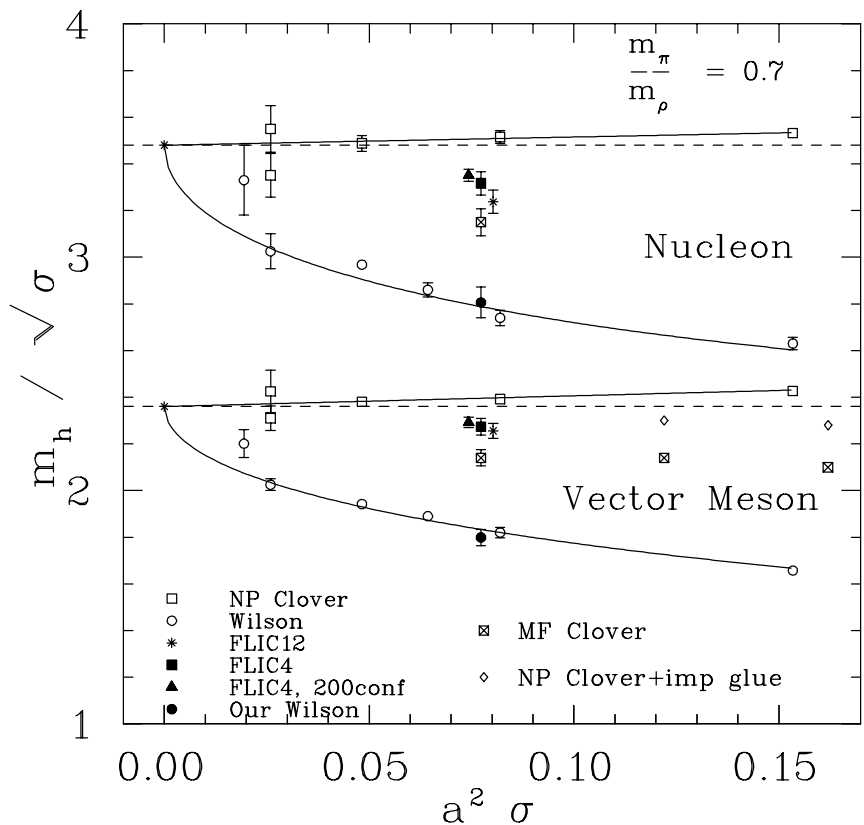} }
\vspace*{0.5cm}
\caption{Nucleon and vector meson masses for the Wilson, NP-improved,
	mean-field clover and FLIC actions.
	Results from the present simulations, indicated by the solid
	points, are obtained by interpolating the results of
	Fig.~{\protect\ref{Mvsmpi2}} to $m_\pi / m_\rho = 0.7$.
	The fat links are constructed with $n=4$ (solid squares) and
	$n=12$ (stars) smearing sweeps at $\alpha = 0.7$.
	The solid triangles are results for the FLIC4 action when 200
	configurations are used in the analysis.
	The FLIC results are offset from the central value for clarity.
        Our MF clover result at $a^2 \sigma \sim 0.08$ lies systematically
        low relative to the FLIC actions.}
\label{scaling1}
\end{center}
\end{figure}

The scaling behavior of the different actions is illustrated in
Fig.~\ref{scaling1}.
The present results for the Wilson action agree with those of
Ref.~\cite{Edwards:1998nh}.
The first feature to observe in Fig.~\ref{scaling1} is that actions with
fat-link irrelevant operators perform extremely well.
For both the vector meson and the nucleon, the FLIC actions perform
significantly better than the mean-field improved clover action.
It is also clear that the FLIC4 action performs systematically better
than the FLIC12.
This suggests that 12 smearing sweeps removes too much short-distance
information from the gauge-field configurations.
On the other hand, 4 sweeps of smearing combined with our
${\cal O} (a^4)$ improved $F_{\mu\nu}$ provides excellent results,
without the fine tuning of $C_{\rm SW}$ in the NP improvement program.

Notice that for the $\rho$ meson, a linear extrapolation of previous
mean-field improved clover results in Fig.~\ref{scaling1} passes through
our mean-field improved clover result at $a^2 \sigma \sim 0.08$ which 
lies systematically low relative to the FLIC actions.
However, a linear extrapolation does not pass through the continuum limit
result, thus confirming the presence of significant ${\cal O}(a)$ errors
in the mean-field improved clover fermion action.
%
%
While there are no NP-improved clover plus improved glue simulation
results at $a^2 \sigma \sim 0.08$, the simulation results that are
available indicate that the fat-link results also compete well with
those obtained with a NP-improved clover fermion action.

Having determined FLIC4 is the preferred action, we have increased the
number of configurations to 200 for this action.
As expected, the error bars are halved and the central values for the
FLIC4 points move to the upper
end of the error bars on the 50 configuration result, further supporting the 
promise of excellent scaling.

%
\begin{figure}[t]
\begin{center}
\epsfysize=11.3truecm
\leavevmode
\rotate[l]{\epsfbox{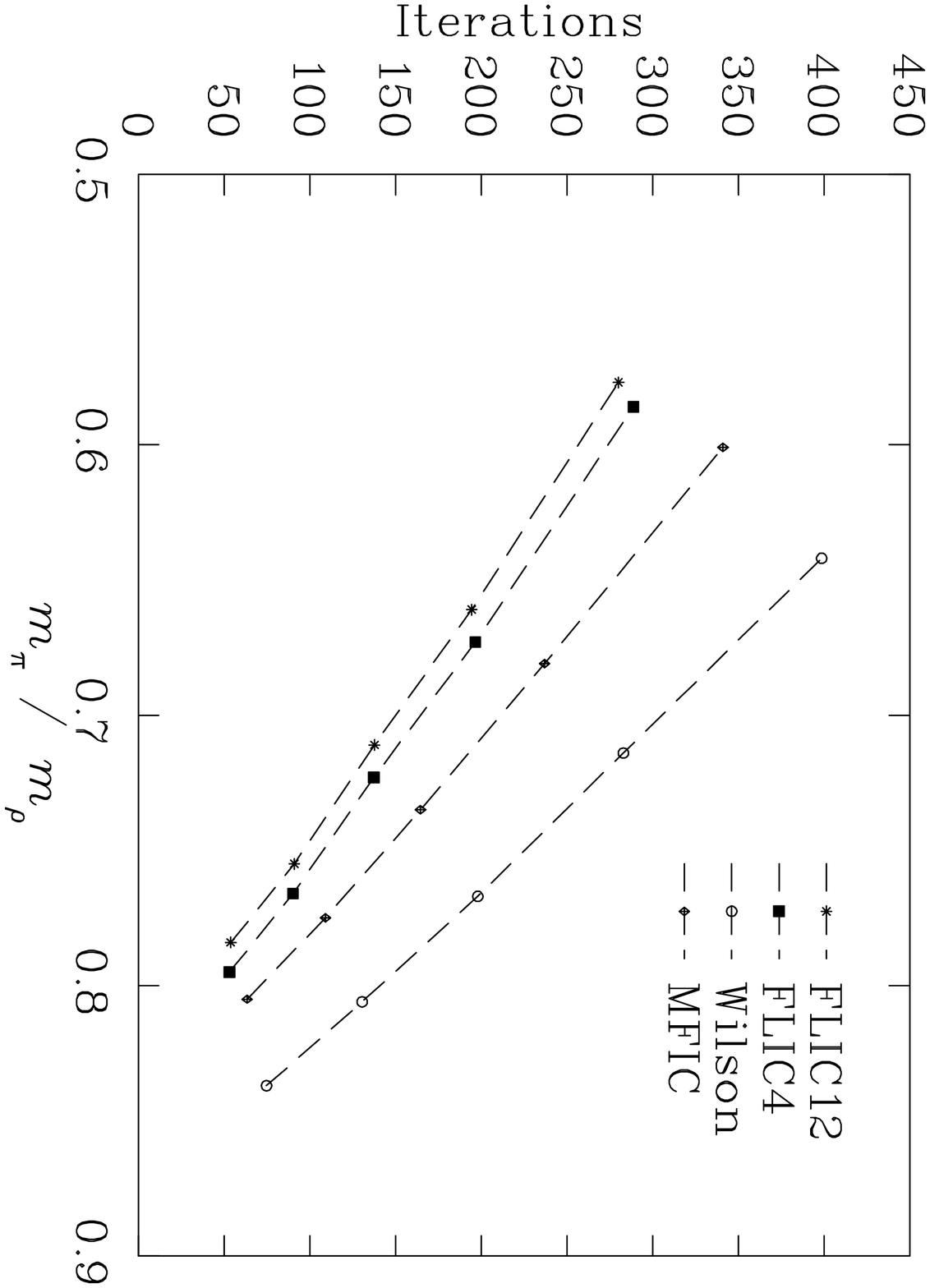} }
\vspace*{0.5cm}
\caption{Average number of stabilized biconjugate gradient iterations for
	the Wilson, FLIC and mean-field improved clover (MFIC) actions
        plotted against $m_{\pi} / m_{\rho}$.
	The fat links are constructed with $n=4$ (solid squares) and
	$n=12$ (stars) smearing sweeps at $\alpha = 0.7$.}
\label{iterations}
\end{center}
\end{figure}

Finally, in Fig.~\ref{iterations} we compare the convergence rates of the
different actions by plotting the number of stabilized biconjugate
gradient \cite{BiCG} iterations required to invert the fermion matrix as
a function of $m_{\pi} / m_{\rho}$.
For any particular value of $m_{\pi} / m_{\rho}$, the FLIC actions
converge faster than both the Wilson and mean-field improved clover
fermion actions.
Also, the slopes of the FLIC lines are smaller in magnitude than those
for Wilson and mean-field improved clover actions, which provides great
promise for performing cost effective simulations at quark masses closer
to the physical values.
Problems with exceptional configurations have prevented such simulations 
in the past.

\section{Conclusions}
\label{conclusion}

We have examined the hadron mass spectrum using a novel Fat-Link
Irrelevant Clover (FLIC) fermion action, in which only the irrelevant,
higher-dimension operators involve smeared links.
One of the main conclusions of this work is that the use of fat links in
the irrelevant operators provides excellent results. Fat links promise
improved scaling behavior over mean-field improvement. This technique
also solves a significant problem with ${\cal O}(a)$ nonperturbative
improvement on mean field-improved gluon configurations. Simulations
are possible and the results are competitive with 
nonperturbative-improved clover results on plaquette-action gluon
configurations.
We have found that minimal smearing holds the promise of better scaling
behavior.
Our results suggest that too much smearing removes relevant information
from the gauge fields, leading to a poorer performance. 
Fermion matrix inversion for FLIC actions is more efficient and results
show no sign of exceptional configuration problems.

This work paves the way for promising future studies.
It will be of great interest to consider different lattice spacings to
further test the scaling of the fat-link actions.
Furthermore, the exceptional configuration issue can be explored by
pushing the quark mass down to lower values.
We are currently performing simulations at $m_{\pi}/m_{\rho}=0.36$ and 
these results will available soon.
%
%
A study of the spectrum of excited hadrons using the FLIC
actions is also currently in progress \cite{Nstar}.

\acknowledgements

This work was supported by the Australian Research Council.
We are grateful to Herbert Neuberger for helpful discussions regarding the 
gauge invariance of APE smearing.
We have also benefited from discussions with Robert Edwards and Urs Heller
during the Lattice Hadron Physics workshop held in Cairns, Australia.
%
We would also like to thank the Australian National Computing Facility
for Lattice Gauge Theories for the use of the Orion Supercomputer.
W.M. and F.X.L. were partially supported by the U.S. Department of Energy
contract \mbox{DE-AC05-84ER40150}, under which the Southeastern
Universities Research Association (SURA) operates the Thomas Jefferson
National Accelerator Facility (Jefferson Lab).


\end{document}